\documentclass[12pt]{article}
\usepackage{amssymb,amsmath}
\usepackage{epsfig}
\def\L{{\mathcal L}}

\def\L{{\mathcal L}}

\def\ds{\displaystyle}

\def\a{\alpha}
\def\b{\beta}

\def\m{\mu}

\def\l{\lambda}

\def\p{\partial}
\def\rb{\right}
\def\lb{\left}

\newcommand{\eq}[1]{\begin{equation} #1 \end{equation}}
\newcommand{\al}[1]{\begin{align} #1 \end{align}}
\newcommand{\ml}[1]{\begin{multline} #1 \end{multline}}

\newcommand{\ba}{\begin{eqnarray}}
\newcommand{\ea}{\end{eqnarray}}
\newcommand{\nl }{ \nonumber  }
\newcommand{\hs}{\hspace{0.5cm}}

\begin{document}
\vspace*{2.cm}
\begin{center}


{\Large \bf On the multi-spin magnon and spike solutions from membranes}

\vspace*{1cm} P. Bozhilov${}^{\star}$\footnote{e-mail:
bozhilov@inrne.bas.bg} and R.C.
Rashkov${}^{\dagger}$\footnote{e-mail: rash@hep.itp.tuwien.ac.at,
on leave from Dept of Physics, Sofia University, Bulgaria.}

\ \\

${}^{\star}$ \textit{Institute for Nuclear Research and Nuclear
Energy, Bulgarian Academy of Sciences, \\ 1784 Sofia, Bulgaria}

${}^{\dagger}$\textit{Institute for Theoretical Physics, Vienna
University of Technology,\\
Wiedner Hauptstr. 8-10, 1040 Vienna, Austria}

\end{center}

\begin{abstract}

Recently important classes of solitonic string solutions were
obtained - giant magnons and single spikes. In previous study we
showed the existence of giant magnon-like membrane solutions and
studied their properties. In this paper we investigate  classical
rotating membranes representing analog of a specific class of
string spiky solutions. Using the reduction to the
Neumann-Rosochatius integrable system we find analog of the string
single spike solutions. In contrast to the magnon-like solutions,
this case is characterized with finite difference of energy and
``winding number'' and finite spins as well.

\end{abstract}

\section{Introduction}

Recent advances indicate that integrable structures play an
important role in relating string/M-theory and the gauge theories.
One of the conjectures of the correspondence is the direct
relation between the spectrum of anomalous dimensions of gauge
invariant operators in the the gauge theory and the energy
spectrum of the string/M-theory. Therefore, one of the important
issues is the dispersion relations in the string/M-theory. The
techniques of integrable systems have therefore become useful in
studying this correspondence in details.

The strong/weak nature of the duality between string/M-theory and
gauge theory makes the proof of the conjectured correspondence
difficult. Basically, when the perturbative study is valid on one
side of the correspondence, one cannot trust the result from the
other, strong coupled theory. There exist however special regimes
where one can do reliable calculations on both side of the
correspondence. Assuming integrability, in the strong-coupling
limit the $S$ matrix can be interpreted as describing the two-body
scattering of elementary excitations on the worldvolume. When
their worldvolume momenta become large, these excitations can be
described as special types of solitonic solutions, or giant
magnons, and the interpolating region is described by the dynamics
of the so-called near-flat-space regime
\cite{Hofman:2006xt,Swanson:2006dec}. On the gauge theory side,
the action of the dilatation operator on single-trace operators is
the same as that of a Hamiltonian acting on the states of a
certain spin chain \cite{Minahan:2002ve}. This turns out to be of
great advantage because one can diagonalize the matrix of
anomalous dimensions by using the algebraic Bethe ansatz technique
(see \cite{fadd} for a nice review on the algebraic Bethe ansatz).
In this picture the dynamics involves diffractionless scattering
encoded in the S matrix. Proving that the gauge and string/M
theories are identical in the planar limit therefore amounts to
showing that the underlying physics of both theories is governed
by the same two-body scattering matrix.

This approach was developed mainly for checking string/gauge
theory duality. It is still not quite well understood in the case
of M-theory/gauge theory correspondence due to the complexity of
the latter. It is worth however to study particular solutions of
the above type for at least three reasons. First, obtaining
solutions of stringy type for membranes is interesting on its own
and is expected to shed light on some properties of the theory.
Secondly, one can gain some insights about how the correspondence
should relate the spectrum to the gauge theory operators and
identify the latter, and finally it should be possible to
reproduce the string result upon dimensional reduction from eleven
to ten dimensions.

Recently Hofman and Maldacena \cite{Hofman:2006xt} were able to
map spin chain "magnon" excitations to specific rotating
semiclassical string states on $R\times S^2$. The relation between
energy and angular momentum for the one spin giant magnon found in
\cite{Hofman:2006xt} is:
\eq{E-J=\ds\frac{\sqrt{\l}}{\pi}|\sin\ds\frac{p}{2}| } where $p$
is the magnon momentum which on the string side is interpreted as
a difference in the angle $\phi$ (see \cite{Hofman:2006xt} for
details). These classical string solutions were further
generalized to include magnon bound states
(\cite{Dorey1},\cite{Dorey2},\cite{AFZ},\cite{MTT}, as well as
dynamics on the whole $S^5$ \cite{volovich},
\cite{Kruczenski:2006pk}  and in fact a method to construct
classical string solutions describing superposition of arbitrary
scattering and bound states was found \cite{Kalousios:2006xy}. The
semiclassical quantization of the giant magnon solution was
performed in \cite{Papathanasiou:2007gd}\footnote{See also
\cite{WHH06}-\cite{BV0707}.}.

A class of classical string solutions with spikes may be of
interest for the AdS/CFT correspondence as well. These spiky
strings were constructed in the $AdS_5$ subspace of $AdS_5\times
S^5$ \cite{Kruczenski:2004wg} and it was argued that they
correspond to single trace operators with a large number of
derivatives. The spiky strings were generalized to include
dynamics in the $S^5$ part of $AdS_5\times S^5$
\cite{Ryang:2005yd} and in \cite{McLoughlin:2006tz} closed strings
with "kinks" were considered. Some generalizations can be found
also in \cite{Chu:2006ae,Bobev:2006fg}.

Leaded by the recent developments in string theory, we would like
to study the existence of so called single spike solutions
\cite{IK0705,Mosaffa:2007,BR0706} in the membrane case. As it is
already known, one can find a consistent reduction of the membrane
to the Neumann-Rosochatius integrable system \cite{MNR} and
therefore one can expect that such solutions can be obtained in
this  case as well.

The paper is organized as follows. In the next section we give
short review of the single spike sector in the string case. To
analyze the problem we use reduction to Neumann-Rosochatius (NR)
integrable system and reproduce some results from
\cite{IK0705,Mosaffa:2007,BR0706}. Next we proceed with the
analysis of the membrane case. We use the same technique, namely
reduction to the NR system, to obtain the corresponding dispersion
relations. We summarize our study in short concluding section.

\section{Spiky solutions - string case}
To obtain single spike solutions we will use the reduction to the
NR system. This approach is convenient because it will be used in
the membrane case and then the comparison and dimensional
reduction become transparent. We start with the Polyakov sigma
model action supplied with the corresponding Visrasoro
constraints.

\paragraph{The Lagrangian} \ \\
To write the Lagrangian of the theory we use the following
parametrization \eq{ x_a(\xi)=r_a(\xi)e^{i\m_a(\xi)} , \quad
|x_a'|^2={r_a'}^2+r_a^2{\m_a'}^2, \quad \xi=\a\sigma+\b\tau.
\label{parametr}} Using parametrization in terms of complex
variables as above one can find the Lagrangian \ml{
\L=\sum\limits_a\lb[(\a^2-\b^2){r_a'}^2+(\a^2-\b^2)r_a^2\lb(\m_a'-\frac{\b\omega_a}{\a^2-\b^2}\rb)^2-
\frac{\a^2}{\a^2-\b^2}\omega_a^2r_a^2\rb]\\
+\Lambda\lb(\sum\limits_a r_a^2-1\rb). \label{lagr-1} }
Integrating once the equations of motion for $\m_a$, we get \eq{
\m_a'=\frac{1}{\a^2-\b^2}\lb(\frac{C_a}{r_a^2}+\b\omega_a\rb)
\label{eq-mu} } where $C_a$ are constants of motion. On other hand
for $r_a$ we get \eq{
(\a^2-\b^2)r_a''-\frac{C_a^2}{\a^2-\b^2}\frac{1}{r_a^3}+\frac{\a^2}{\a^2-\b^2}\omega_a^2r_a-\Lambda
r_a=0 \label{eq-r} } where we used the equations for $\mu$
(\ref{eq-mu}).

The  equations for $r_a$ can be then obtained from the effective
Lagrangian \eq{
\L=\sum\limits_a\lb[(\a^2-\b^2){r_a'}^2-\frac{C_a^2}{\a^2-\b^2}\frac{1}{r_a^2}-\frac{\a^2}{\a^2-\b^2}\omega_a^2r_a^2\rb]
+\Lambda\lb(\sum\limits_a r_a^2-1\rb). \label{lagr-nr} } This few
steps were just to make link to string case where the
corresponding solutions  are already known. One can easily find
the corresponding Hamiltonian \eq{ \mathcal H=
\sum\limits_a\lb[(\a^2-\b^2){r_a'}^2+\frac{C_a^2}{\a^2-\b^2}\frac{1}{r_a^2}
+\frac{\a^2}{\a^2-\b^2}\omega_a^2r_a^2\rb]. \label{hamil-1} }

\paragraph{Conformal constraints in NR} \ \\
An important part of the analysis are Virasoro constraints \al{
& \sum\limits_a\lb[|\p_\tau X_a|^2+|\p_\sigma X_a|^2\rb]=\kappa^2 \label{vir1} \\
& \sum\limits_a\lb[ \p_\tau X_a\p_\sigma\bar X_a+\p_\tau\bar
X_a\p_\sigma X_a\rb]=0 \label{vir2} } where we used \eq{
X_a=x_a(\xi)e^{i\omega_a\tau}, \quad \xi=\a\sigma+\b\tau, \quad
x_a(\xi+2\pi\a)=x_a(\xi) \label{period-x} } One can use the ansatz
(\ref{parametr}) and the solutions (\ref{eq-mu}) to find
constraints in terms of the parameters. Namely, one can write  the
final form of the Virasoro constraints as \al{
& H=\frac{\a^2+\b^2}{\a^2-\b^2}\kappa^2 \label{vir1-fin-nr} \\
& \sum\limits_a\omega_aC_a+\b\kappa^2=0 \label{vir2-fin-nr} }

The periodicity conditions on $r_a$ and $\m_a$ should be imposed
and they are as follows \eq{ r_a(\xi+2\pi\a)=r_a(\xi), \quad
\mu_a(\xi+2\pi\a)=\mu_a(\xi)+2\pi n_a, \,\, n_a - \text{integer} }

\paragraph{Two-spin spiky solutions} \ \\
What we have to do is to satisfy the constraints
(\ref{vir1-fin-nr}), (\ref{vir2-fin-nr}) in a way that we get
single spike solutions. The two-spin case means that we reduce the
system to the $S^3$ subspace, or  $r_3=\m_3=0$ (also $C_3=0$). In
addition, the condition that one of the turning points of the
string is at $\theta=\pi/2$ imposes $ C_2=0$,
$\b=-\frac{\omega_1C_1}{\kappa^2}$ (we set $\a=1$ and then $|\b|$
becomes a ''group velocity``). The equation for $\kappa$ has two
solutions: $\kappa=\omega_1$ corresponding to a magnon type
solution, and $\kappa= C_1$ corresponding to single spike
solution.

The single spike choice for $\kappa$ and other parameters
is\footnote{For details see \cite{IK0705,Mosaffa:2007,BR0706}}
\eq{ \kappa=C_1, \quad
\beta=-\frac{\omega_1C_1}{\kappa^2}=-\frac{\omega_1}{C_1} } We
have to solve the following equation \eq{
\frac{d\,u}{d\xi}=u'=2\frac{\sqrt{\Delta\omega^2}}{1-\beta^2}(1-u)\sqrt{u-\bar
u} } where \al{ & u=r_1^2=sin^2\theta, \quad \bar
u=\frac{C_1^2}{\sqrt{\Delta\omega^2}}, \quad
 \Delta\omega^2 = \omega_1^2-\omega_2^2,  \notag \\
&
d\xi=\frac{du}{u'}=\frac{(1-\beta^2)\,du}{2\sqrt{\Delta\omega^2}(1-u)\sqrt{u-\bar
u}}= \frac{(C_1^2-\omega_1^2)\,d
u}{2C_1^2\sqrt{\Delta\omega^2}(1-u)\sqrt{u-\bar u}} }

The conserved quantities are \al{
& \hat\m_1=\frac{C_1}{(1-\beta^2)}\int\frac{d\xi}{u} +\frac{\beta\omega_1}{(1-\beta^2)}\int d\xi \notag \\
& E=\kappa T\int d\xi \notag \\
& J_1=\frac{C_1\beta}{(1-\beta^2)}\int d\xi + \frac{\omega_1}{(1-\beta^2)}T\int u\,d\xi \\
& J_2=\frac{\omega_2}{(1-\beta^2)}T\int(1-u)d\xi \notag } where
$\hat\m_1$ is the angular extension of the string \ba\nl
\hat\m_1=\int \mu'_1 d\xi.\ea

Let us consider the following difference \eq{
E-T\hat\m_1=\frac{2C_1T}{\sqrt{\Delta\omega^2}}\arccos\sqrt{\bar
u}=\frac{\sqrt{\lambda}}{\pi}\bar\theta } where \eq{
\bar\theta=\frac{\pi}{2}-\theta_0. } Unlike in the case of giant
magnons, this difference is finite (and the energy is still
large). This means that the string has one spike, but is wounded
many times along the equator of $S^5$ so that the difference
between the energy and winding number $T\hat\m_1$ is finite.

For the spin $J_1$ we get \eq{
J_1=\frac{2T\omega_1}{\sqrt{\Delta\omega^2}}\cos\theta_0=\frac{2T\omega_1}{\sqrt{\Delta\omega^2}}\sin\bar\theta.
} and analogously for $J_2$ we find \eq{
J_2=-\frac{2T\omega_2}{\sqrt{\Delta\omega^2}}\cos\theta_0=-\frac{2T\omega_2}{\sqrt{\Delta\omega^2}}\sin\bar\theta
} Defining $\sin\gamma$ as in \cite{IK0705}, equation (6.23)
(where it is actually wrong because $\omega_1>\omega_2$) \eq{
\sin\gamma=\frac{\omega_2}{\omega_1},\hs
\sin\theta_0=\frac{C_1}{\sqrt{\Delta\omega^2}} } we one can write
$J_i$ as \al{
& J_1=2T\frac{1}{\sin\gamma}\sin\bar\theta \notag \\
& J_2=-2T\frac{\sin\gamma}{\cos\gamma}\sin\bar\theta } Note that
both $J_i$ are finite.

Eliminating the auxiliary parameter $\gamma$ one can obtain the
following dispersion relations \al{
& E-T\hat\m_1=\frac{\sqrt{\lambda}}{\pi}\bar\theta \label{ssse} \\
& J_1=\sqrt{J_2^2+\frac{\lambda}{\pi^2}\sin^2\bar\theta}.
\label{sssj}} We see that in the limit $J_2=0$  we reproduce the
one-spin case. One can define the anomalous dimensions in this
case ($J_2=0$ and $J_1=J$) as in the magnon considerations but
subtracting the winding number $T\hat\m_1$ to get ''renormalized``
result \eq{
\Delta=(E-T\hat\m_1)-J=\frac{\sqrt{\lambda}}{\pi}\left(\bar\theta-\sin\bar\theta\right),
} which reproduces the result of
\cite{IK0705,Mosaffa:2007,BR0706}.

\setcounter{equation}{0}
\section{Membranes on $AdS_4\times S^7$}
Turning to the membrane case, let us first write down the gauge
fixed membrane action and constraints in diagonal worldvolume
gauge, we are going to work with: \ba\label{omagf} &&S_{M}=\int
d^{3}\xi \mathcal{L}_{M}= \int
d^{3}\xi\left\{\frac{1}{4\lambda^0}\Bigl[G_{00}-\left(2\lambda^0T_2\right)^2\det
G_{ij}\Bigr] + T_2 C_{012}\right\},
\\ \label{00gf} &&G_{00}+\left(2\lambda^0T_2\right)^2\det G_{ij}=0,
\\ \label{0igf} &&G_{0i}=0.\ea They {\it coincide} with the
frequently used gauge fixed Polyakov type action and constraints
after the identification $2\lambda^0T_2=L=const$, where
$\lambda^0$ is Lagrange multiplier and $T_2$ is the membrane
tension. In (\ref{omagf})-(\ref{0igf}), the fields induced on the
membrane worldvolume $G_{mn}$ and $C_{012}$ are given by
\ba\label{im} &&G_{mn}= g_{MN}\p_m X^M\p_n X^N,\hs C_{012}=
c_{MNP}\p_{0}X^{M}\p_{1}X^{N}\p_{2}X^{P}, \\ \nl
&&\p_m=\p/\p\xi^m,\hs m = (0,i) = (0,1,2),\\ \nl
&&(\xi^0,\xi^1,\xi^2)=(\tau,\sigma_1,\sigma_2),\hs M =
(0,1,\ldots,10),\ea where $g_{MN}$ and $c_{MNP}$ are the
components of the target space metric and 3-form gauge field
respectively.

Searching for membrane solutions in $AdS_4\times S^7$ analogous to
string ones on $AdS_5\times S^5$, we should first eliminate the
membrane interaction with the background 3-form field on $AdS_4$.
To make our choice, let us write down the background. It can be
parameterized as follows \ba\nl
&&ds^2=(2l_p\mathcal{R})^2\left[-\cosh^2\rho
dt^2+d\rho^2+\sinh^2\rho\left(d\alpha^2+\sin^2\alpha
d\beta^2\right) + 4d\Omega_7^2\right], \\ \nl
&&c_{(3)}=(2l_p\mathcal{R})^3\sinh^3\rho\sin\alpha dt\wedge
d\alpha\wedge d\beta.\ea Since we want the membrane to have
nonzero conserved energy and spin on $AdS$, the choice for which
the interaction with the $c_{(3)}$ field disappears is\footnote{Of
course, we can fix the angle $\beta$ instead of $\alpha$. We
choose to fix $\alpha$ because $\beta$ is one of the isometry
coordinates in the initial $AdS_4$ space.}: \ba\nl
\alpha=\alpha_0=const.\ea The metric of the corresponding subspace
of $AdS_4$ is \ba\label{sub}
&&ds^2_{sub}=(2l_p\mathcal{R})^2\left(-\cosh^2\rho
dt^2+d\rho^2+\sinh^2\rho\sin^2\alpha_0 d\beta^2\right)=
\\ \nl &&(2l_p\mathcal{R})^2\left[-\cosh^2\rho
dt^2+d\rho^2+\sinh^2\rho d(\beta\sin\alpha_0)^2\right].\ea Hence,
the appropriate membrane embedding into (\ref{sub}) and $S^7$,
analogous to the string embedding in $AdS_5\times S^5$, is \ba\nl
&&Z_{\mu}=2l_p\mathcal{R}\mbox{r}_\mu(\xi^m)e^{i\phi_\mu(\xi^m)},\hs
\mu=(0,1),\hs \phi_{\mu}=(\phi_0,\phi_1)=(t,\beta\sin\alpha_0),\\
\label{gme}
&&\hs\hs\hs\hs\hs\hs\hs\hs\hs\hs \eta^{\mu\nu} \mbox{r}_\mu\mbox{r}_\nu+1=0,\hs \eta^{\mu\nu}=(-1,1), \\
\nl &&W_{a}=4l_p\mathcal{R}r_a(\xi^m)e^{i\varphi_a(\xi^m)},\hs
a=(1,2,3,4),\hs \delta_{ab} r_ar_b-1=0.\ea For this embedding, the
induced metric is given by \ba\label{mim}
&&G_{mn}=\eta^{\mu\nu}\p_{(m}Z_\mu\p_{n)}\bar{Z_\nu} +
\delta_{ab}\p_{(m}W_a\p_{n)}\bar{W_b}= \\ \nl
&&(2l_p\mathcal{R})^2\left[\sum_{\mu,\nu=0}^{1}\eta^{\mu\nu}\left(\p_m\mbox{r}_\mu\p_n\mbox{r}_\nu
+ \mbox{r}_\mu^2\p_m\phi_\mu\p_n\phi_\nu\right) +
4\sum_{a=1}^{4}\left(\p_mr_a\p_nr_a +
r_a^2\p_m\varphi_a\p_n\varphi_a\right)\right].\ea We will use the
expression (\ref{mim}) for $G_{mn}$ in (\ref{omagf}), (\ref{00gf})
and (\ref{0igf}). Correspondingly, the membrane Lagrangian becomes
\ba\label{geml}
\mathcal{L}=\mathcal{L}_{M}+\Lambda_A(\eta^{\mu\nu}
\mbox{r}_\mu\mbox{r}_\nu+1)+\Lambda_S(\delta_{ab} r_ar_b-1),\ea
where $\Lambda_A$ and $\Lambda_S$ are Lagrange multipliers.

In this paper, we are interested in the following particular case
of the membrane embedding (\ref{gme}) \ba\label{rme}
Z_{0}=2l_p\mathcal{R}e^{i\kappa\tau},\hs Z_1=0,\ea which implies
\ba\nl \mbox{r}_0=1,\hs \mbox{r}_1=0,\hs \phi_0=t=\kappa\tau.\ea
For this ansatz, the metric induced on the membrane worldvolume
simplifies to \ba\label{mimpc} G_{mn}=(4l_p\mathcal{R})^2\left[
\sum_{a=1}^{4}\left(\p_mr_a\p_nr_a +
r_a^2\p_m\varphi_a\p_n\varphi_a\right)-\delta_{m}^{0}\delta_{n}^{0}(\kappa/2)^2\right],\ea
and the membrane Lagrangian is given by \ba\label{pcml}
\mathcal{L}=\mathcal{L}_{M}+\Lambda_S\left(\sum_{a=1}^{4}r_a^2-1\right).\ea

The known most general membrane embedding in $AdS_4\times S^7$
leading to the Neumann - Rosochatius integrable system is
\cite{MNR} \ba\label{ome}
&&Z_{0}=2l_p\mathcal{R}e^{i\kappa\tau},\hs Z_1=0,\hs
W_{a}=4l_p\mathcal{R}r_a(\xi,\eta)e^{i\left[\omega_{a}\tau+g_a(\xi,\eta)\right]},\\
\nl &&\xi=\alpha\sigma_1+\beta\tau,\hs
\eta=\gamma\sigma_2+\delta\tau,\hs \alpha,\beta,\gamma,\delta =
constants,\ea for \ba\nl &&r_{1}=r_{1}(\xi),\hs
r_{2}=r_{2}(\xi),\hs \omega_3=\pm\omega_4=\omega,\\ \label{sol}
&&r_3=r_3(\eta)=a\sin(b\eta+c),\hs
r_4=r_4(\eta)=a\cos(b\eta+c),\hs a<1, \\ \nl &&g_1=g_1(\xi),\hs
g_2=g_2(\xi),\hs a,b,c,g_3,g_4=constants,\hs \delta=0.\ea The
above ansatz reduces the membrane Lagrangian $\mathcal{L}$ in
(\ref{pcml}) to \ba\nl
\mathcal{L}_*^M&=&-\frac{(2l_p\mathcal{R})^2}{\lambda^0}
\left\{\sum_{a=1}^{2}\left[(\tilde{A}^2-\beta^2)(\p_\xi r_a)^2 +
(\tilde{A}^2-\beta^2)r_a^2\left(\p_\xi
g_a-\frac{\beta\omega_a}{\tilde{A}^2-\beta^2}\right)^2\right.\right.\\
\label{NRml} &-&
\left.\left.\frac{\tilde{A}^2}{\tilde{A}^2-\beta^2}\omega_a^2
r_a^2\right] + (\kappa/2)^2-(a\omega)^2\right\} +
\Lambda_S\left[\sum_{a=1}^{2}r_a^2-(1-a^2)\right],\ea where \ba\nl
\tilde{A}^2\equiv
\left(8\lambda^0T_2l_p\mathcal{R}ab\alpha\gamma\right)^2.\ea As
shown in \cite{MNR}, $\mathcal{L}_*^M$ corresponds to the
following Neumann-Rosochatius type Lagrangian \ba\nl
\mathcal{L}_{NR}^M&=&\frac{(2l_p\mathcal{R})^2}{\lambda^0}
\sum_{a=1}^{2}\left[(\tilde{A}^2-\beta^2)(\p_\xi r_a)^2 -
\frac{C_a^2}{(\tilde{A}^2-\beta^2)r_a^2}
- \frac{\tilde{A}^2}{\tilde{A}^2-\beta^2}\omega_a^2 r_a^2\right] \\
\nl &+&\Lambda_S\left[\sum_{a=1}^{2}r_a^2-(1-a^2)\right],\hs
C_a=constants.\ea

For the present case, the constraint $G_{02}=0$ in (\ref{0igf}) is
satisfied identically, due to $\delta=0$. The other two
constraints in (\ref{00gf}) and (\ref{0igf}) can be written in the
form \ba\nl H &\sim &
\sum_{a=1}^{2}\left[(\tilde{A}^2-\beta^2)(\p_\xi r_a)^2 +
\frac{C_a^2}{(\tilde{A}^2-\beta^2)r_a^2}
+ \frac{\tilde{A}^2}{\tilde{A}^2-\beta^2}\omega_a^2 r_a^2\right]\\
\label{MNR00}
&&=\frac{\tilde{A}^2+\beta^2}{\tilde{A}^2-\beta^2}\left[(\kappa/2)^2-(a\omega)^2\right],
\\ \label{MNR01} &&\sum_{a=1}^{2}\omega_{a}C_a +
\beta\left[(\kappa/2)^2-(a\omega)^2\right]=0,\ea where $r_a$ must
satisfy the condition \ba\label{c}
\sum_{a=1}^{2}r_a^2-(1-a^2)=0.\ea Parameterizing the circle
(\ref{c}) by \ba\nl r_1=(1-a^2)^{1/2}\cos\psi,\hs
r_2=(1-a^2)^{1/2}\sin\psi,\ea one obtains that (\ref{c}) is
satisfied identically and (\ref{MNR00}) reduces to (prime is used
for $d/d\xi$) \ba\label{pps}
\psi'^2&=&\frac{1}{(\tilde{A}^2-\beta^2)^2(1-a^2)}\left\{(\tilde{A}^2+\beta^2)
\left[(\kappa/2)^2-(a\omega)^2\right]\right.\\ \nl
&&-\left.\frac{1}{1-a^2}\left(\frac{C_1^2}{\cos^2\psi}+\frac{C_2^2}{\sin^2\psi}\right)
-\tilde{A}^2(1-a^2)(\omega_1^2\cos^2\psi +
\omega_2^2\sin^2\psi)\right\}.\ea

\paragraph{One-spin solutions}\ \\
If we set $C_1=\omega_1=0$, from (\ref{MNR01}) and (\ref{pps}) we
get \ba\label{1sc}
\psi'=\pm\frac{\tilde{A}\omega_2}{(\tilde{A}^2-\beta^2)\sin\psi}
\sqrt{(\sin^2\psi_0-\sin^2\psi)(\sin^2\psi-\sin^2\psi_1)},\ea
where \ba\nl
\sin^2\psi_0=\frac{(\kappa/2)^2-(a\omega)^2}{\omega_2^2(1-a^2)},\hs
\sin^2\psi_1= \frac{\beta^2}{\tilde{A}^2}\sin^2\psi_0.\ea The
equation (\ref{1sc}) gives the same spiky and giant magnon
solutions as (5.11) of \cite{IK0705}.

\paragraph{Two-spin solutions}\ \\
Following \cite{BR0706}, for obtaining two-spin, spiky and giant
magnon solutions, we set $C_1=0$ in (\ref{pps}) and taking into
account (\ref{MNR01}), we receive \ba\nl
\psi'^2&=&\frac{\tilde{A}^2(\omega_2^2-\omega_1^2)}{(\tilde{A}^2-\beta^2)^2\sin^2\psi}\left\{
\frac{(\kappa/2)^2-(a\omega)^2}{(1-a^2)(\omega_2^2-\omega_1^2)}
\left[1+\frac{\beta^2}{\tilde{A}^2}-\frac{(1-a^2)\omega_1^2}{(\kappa/2)^2-(a\omega)^2}\right]\sin^2\psi\right.\\
\label{ppt}
&&-\left.\frac{\beta^2\left[(\kappa/2)^2-(a\omega)^2\right]^2}{\tilde{A}^2(1-a^2)^2\omega_2^2(\omega_2^2-\omega_1^2)}
-\sin^4\psi\right\}.\ea The solutions we are searching for,
correspond to two particular choices of the parameters in the
above expression for $\psi'^2$.

The first choice is to set in (\ref{ppt}) \ba\nl
(\kappa/2)^2-(a\omega)^2=(1-a^2)\omega_2^2.\ea This results in
\ba\label{2sgm}
\psi'=\frac{\tilde{A}\sqrt{\omega_2^2-\omega_1^2}}{\tilde{A}^2-\beta^2}\frac{\cos\psi}{\sin\psi}
\sqrt{\sin^2\psi-\sin^2\psi_0},\ea where \ba\nl
\sin^2\psi_0=\frac{\beta^2\omega_2^2}{\tilde{A}^2(\omega_2^2-\omega_1^2)}.\ea
Comparing (\ref{2sgm}) with (3.23) in \cite{BR0706}, we see that
this equation for $\psi$ leads to giant magnon solution on $S^3$.

The second appropriate choice is \ba\nl
(\kappa/2)^2-(a\omega)^2=\frac{\tilde{A}^2(1-a^2)\omega_2^2}{\beta^2}.\ea
Then, (\ref{ppt}) reduces to \ba\label{2sss}
\psi'=\frac{\tilde{A}\sqrt{\omega_2^2-\omega_1^2}}{\tilde{A}^2-\beta^2}\frac{\cos\psi}{\sin\psi}
\sqrt{\sin^2\psi-\sin^2\psi_1},\ea where \ba\nl
\sin^2\psi_1=\frac{\tilde{A}^2\omega_2^2}{\beta^2(\omega_2^2-\omega_1^2)}.\ea
Comparing (\ref{2sss}) with (3.28) in \cite{BR0706}, we see that
this equation for $\psi$ leads to single spike solution on $S^3$,
first obtained in \cite{IK0705} from the Nambu-Goto string action.

Let us finally note that both equations (\ref{2sgm}) and
(\ref{2sss}) are of the type \ba\nl
\frac{d\,u}{d\xi}=u'=2\tilde{K}(1-u)\sqrt{u-\bar u},\hs
\tilde{K}=\frac{\tilde{A}\sqrt{\omega_2^2-\omega_1^2}}{\tilde{A}^2-\beta^2},
\hs u=\sin^2\psi,\ea corresponding to different values of the
parameter $\bar{u}$: \ba\nl \bar u=\sin^2\psi_0,\hs\mbox{or}\hs
\bar u=\sin^2\psi_1.\ea

\paragraph{Conserved quantities}\ \\
The energy $E$ and the angular momenta $J_a$ can be computed by
using the equalities \ba\nl E=-\int
d^2\sigma\frac{\p\mathcal{L}}{\p(\p_0t)},\hs J_a=\int
d^2\sigma\frac{\p\mathcal{L}}{\p(\p_0\varphi_a)},\hs a=1,2,3,4.\ea

For the embedding (\ref{ome}), (\ref{sol}), they are given by
\ba\nl
&&E=\frac{\pi(2l_p\mathcal{R})^2\kappa}{\lambda^0\alpha}\int
d\xi,\\ \nl
&&J_a=\frac{\pi(4l_p\mathcal{R})^2}{\lambda^0\alpha(\tilde{A}^2-\beta^2)}\int
d\xi \left(\beta C_a + \tilde{A}^2\omega_a r_a^2\right),\hs
a=1,2.\ea In order to reproduce the string case, we set
$\omega=0$, which leads to $J_3=J_4=0$.

On the solution for $\xi(u)$, $E$ and $J_a$ take the form \ba\nl
&&E=\frac{\pi(2l_p\mathcal{R})^2\kappa}{\lambda^0\alpha\tilde{K}}\int_{\bar{u}}^{1}
\frac{du}{(1-u)\sqrt{u-\bar{u}}},\\ \label{ccsol}
&&J_1=\frac{\pi(4l_p\mathcal{R})^2\tilde{A}^2(1-a^2)\omega_1}{\lambda^0\alpha\tilde{K}(\tilde{A}^2-\beta^2)}
\int_{\bar{u}}^{1} \frac{du}{\sqrt{u-\bar{u}}},\\ \nl
&&J_2=\frac{\pi(4l_p\mathcal{R})^2}{\lambda^0\alpha\tilde{K}(\tilde{A}^2-\beta^2)}
\left\{\left[\tilde{A}^2(1-a^2)\omega_2-\frac{\beta^2}{\omega_2}(\kappa/2)^2\right]\int_{\bar{u}}^{1}
\frac{du}{(1-u)\sqrt{u-\bar{u}}}\right. \\
\nl &&-\left.\tilde{A}^2(1-a^2)\omega_2\int_{\bar{u}}^{1}
\frac{du}{\sqrt{u-\bar{u}}}\right\}.\ea

From (\ref{ccsol}), for the giant magnon case, one obtains the
energy-charge relation \ba\nl E-\frac{J_2}{2\sqrt{1-a^2}}=
\sqrt{\left(\frac{J_1}{2\sqrt{1-a^2}}\right)^2 + \left[2^7\pi
T_2\left(l_p\mathcal{R}\right)^3a\sqrt{1-a^2}b\gamma\right]^2\cos^2\psi_0}.\ea
The origin of the denominators $2\sqrt{1-a^2}$ in the above
equality can be explained as follows. The factor of $2$ is due to
the different radii of $AdS_4$ and $S^7$, while the radii of
$AdS_5$ and $S^5$ in type IIB string background $AdS_5\times S^5$
are equal. $\sqrt{1-a^2}$ arises because the NR system obtained
from membrane is restricted to lie on a circle with radius
$\sqrt{1-a^2}$, whereas in the string case this radius equals one.

For the single spike solution using (\ref{ccsol}) again, one
arrives at \ba\label{ssj} J_2= \sqrt{J_1^2 + \left[2^8\pi
T_2\left(l_p\mathcal{R}\right)^3a(1-a^2)b\gamma\right]^2\cos^2\psi_1}.\ea

Now following \cite{Kruczenski:2006pk}, we introduce the angular
extension of the membrane as \ba\nl \hat{g}_2=\int g'_2 d\xi =
\frac{1}{\tilde{A}^2-\beta^2}\int\left(\frac{C_2}{r_2^2}
+\beta\omega_2\right)d\xi\ea and compute the difference
$E-\tilde{T}\hat{g}_2$. It turns out it is finite for \ba\nl
\tilde{T}=2^6\pi
T_2\left(l_p\mathcal{R}\right)^3a\sqrt{1-a^2}b\gamma\ea and is
given by \ba\label{sse}
E-\tilde{T}\hat{g}_2=2\tilde{T}\left(\frac{\pi}{2}-\psi_1\right).\ea
Comparing (\ref{ssj}), (\ref{sse}) with (\ref{sssj}),
(\ref{ssse}), we see that the single spike dispersion relations
obtained from membrane and from string are of the same type.

\section{Conclusions}
It is well known that particular solitonic solutions in string/M
theory are important not only on their own but also because of
their relations to the gauge theories and for integrability issues
as well. The development in obtaining such solutions in string
theory inspires the interest to analogous investigations in M
theory. In this short letter we investigated the issue of
existence of spike-like solutions in membrane case. These
solutions extend the class of previously found ''giant magnon``
solutions at the large energy scale. Concretely, these issues are
interesting for the following reasons. The results for type IIB
strings should be reproduced taking an appropriate dimensional
reduction. Then the question whether the known string solutions
have analog in eleven dimensional space-time appears. The answer
to this question for the giant magnon string solutions is positive
\cite{bo-ra:0607}. It is natural to extend the analysis to the
case of so called ''single spike`` solutions.

First we presented a short review of the string case and
demonstrated how one can find solutions of this type using the
reduction to the Neumann-Rosochatius integrable system. In the
next Section we turn to the membrane case. The analysis shows that
there is an analog of this class solutions with the same
properties. To show that, we used again the possibility to  reduce
the theory to a Neumann-Rosochatius integrable system. The
solutions are obtained by taking an appropriate limit and the
dispersion relations are found to be \ba\nl &&\sqrt{1-a^2}
E-\frac{J_2}{2}=\sqrt{\left(\frac{J_1}{2}\right)^2 +
\frac{\tilde{\lambda}}{\pi^2}\sin^2\frac{p}{2}},\\ \nl
&&\tilde{\lambda}=\left[2^7\pi^2T_2\left(l_p\mathcal{R}\right)^3a(1-a^2)b\gamma
\right]^2,\hs \frac{p}{2}=\frac{\pi}{2}- \psi_0,\ea for the giant
magnon solution, and \ba\nl
&&E-\tilde{T}\hat{g}_2=\frac{\sqrt{\tilde{\lambda}}}{\pi}\bar{\psi},
\\ \nl &&\frac{J_2}{2}=\sqrt{\left(\frac{J_1}{2}\right)^2 +
\frac{\tilde{\lambda}}{\pi^2}\sin^2\bar{\psi}},\\ \nl
&&\tilde{T}=\frac{\sqrt{\tilde{\lambda}}}{2\pi},\hs
\bar{\psi}=\frac{\pi}{2}-\psi_1, \ea for the single spike
solution, in full analogy with the string case.

Although the membrane/gauge theory correspondence is not well
developed, we hope that these results will shed light on the
duality. The main obstacle is the lack of clear principle and
mechanism to identify the gauge theory operators.  Collecting
evidences and analogies with strings will show up hopefully such a
principle. There are hopes that one can find an integrable spin
chain which governs this correspondence at least in the sectors
that are known to be integrable but still a lot of work is ahead.

\vspace*{.5cm}
\leftline{\bf Acknowledgements}
\smallskip
This work was supported in part by Bulgarian NSF grants
$F-1412/04$ and $VU-F-201/06$. R.R. acknowledges a Guest
Professorship and warm hospitality at the Institute for
Theoretical Physics, Vienna University of Technology. Many thanks
to Max Kreuzer and his group for fruitful and stimulating
atmosphere. The work of R.R is partially supported by Austrian
Research Fund FWF grant \# P19051-N16.



\begin{thebibliography}{}
\bibitem{Hofman:2006xt}
Diego M. Hofman, Juan Maldacena, \textit{Giant Magnons},  J. Phys.
A 39 (2006) 13095-13118 [arXiv:hep-th/0604135v2].
\bibitem{Swanson:2006dec} Juan Maldacena, Ian Swanson, {\it Connecting giant magnons to the pp-wave:
An interpolating limit of $AdS_5 \times S^5$},
arXiv:hep-th/0612079v4.
\bibitem{Minahan:2002ve} J. A. Minahan and K. Zarembo, \textit{The Bethe-ansatz for
$\mathcal{N}$ = 4 super Yang-Mills}, JHEP 0303 (2003) 013,
[arXiv:hep-th/0212208v3].
\bibitem{fadd} L.~D.~Faddeev, {\it How Algebraic Bethe Ansatz works for integrable model},
arXiv:hep-th/9605187v1.
\bibitem{Dorey1} Nick Dorey, \textit{Magnon bound states and the AdS/CFT
correspondence}, J. Phys. A 39 (2006) 13119-13128,
[arXiv:hep-th/0604175v2].
\bibitem{Dorey2} Heng-Yu Chen, Nick Dorey, Keisuke Okamura
, {\it Dyonic Giant Magnons}, JHEP 0609 (2006) 024,
[arXiv:hep-th/0605155v2].
\bibitem{AFZ} Gleb Arutyunov, Sergey Frolov, Marija Zamaklar,
\textit{Finite-size Effects from Giant Magnons},
doi:10.1016/j.nuclphysb.2006.12.026 [arXiv:hep-th/0606126v2].
\bibitem{MTT} J.A. Minahan, A. Tirziu, A.A. Tseytlin,
\textit{Infinite spin limit of semiclassical string states}, JHEP
0608 (2006) 049, [arXiv:hep-th/0606145v2].
\bibitem{volovich} Marcus Spradlin, Anastasia Volovich, \textit{Dressing the Giant Magnon}, JHEP 0610 (2006) 012
, [arXiv:hep-th/0607009v3].
\bibitem{Kruczenski:2006pk} M. Kruczenski, J. Russo, A.A. Tseytlin, {\it Spiky strings and giant magnons on
$S^5$}, JHEP 0610 (2006) 002 [arXiv:hep-th/0607044v3].
\bibitem{Kalousios:2006xy} Chrysostomos Kalousios, Marcus
Spradlin, Anastasia Volovich, {\it Dressing the Giant Magnon II},
arXiv:hep-th/0611033v1.
\bibitem{Papathanasiou:2007gd} Georgios Papathanasiou, Marcus Spradlin,
{\it Semiclassical Quantization of the Giant Magnon},
arXiv:0704.2389v2 [hep-th].
\bibitem{WHH06} Wung-Hong Huang, {\it Giant Magnons under NS-NS and Melvin
Fields}, JHEP 0612 (2006) 040, [arXiv:hep-th/0607161v4].
\bibitem{OS} Keisuke Okamura, Ryo Suzuki, {\it A Perspective on Classical Strings from
Complex Sine-Gordon Solitons}, Phys. Rev. {\bf D 75} (2007) 046001
[arXiv:hep-th/0609026v3].
\bibitem{H} Shinji Hirano, {\it Fat Magnon}, arXiv:hep-th/0610027v4.
\bibitem{R} Shijong Ryang, {\it Three-Spin Giant Magnons in
$AdS_5\times S^5$}, JHEP 0612 (2006) 043 [arXiv:hep-th/0610037v1].
\bibitem{CDO} Heng-Yu Chen, Nick Dorey, Keisuke Okamura,
{\it The Asymptotic Spectrum of the N=4 Super Yang-Mills Spin
Chain}, arXiv:hep-th/0610295v1.
\bibitem{B06} P. Bozhilov, {\it A note on two-spin magnon-like energy-charge relations
from M-theory viewpoint}, arXiv:hep-th/0612175v2.
\bibitem{M07} J. A. Minahan, {\it  Zero modes for the giant
magnon}, JHEP 0702 (2007) 048 [arXiv:hep-th/0701005v3].
\bibitem{AFGS07} Davide Astolfi, Valentina Forini, Gianluca Grignani, Gordon W.
Semenoff, {\it Gauge invariant finite size spectrum of the giant
magnon}, arXiv:hep-th/0702043v3.
\bibitem{V07} Benoit Vicedo, {\it Giant Magnons and Singular Curves},
arXiv:hep-th/0703180v1.
\bibitem{KNP07} J. Kluson, Rashmi R. Nayak, Kamal L. Panigrahi,
{\it Giant Magnon in NS5-brane Background},
arXiv:hep-th/0703244v2.
\bibitem{Y0704}  C. A. S. Young, {\it q-Deformed Supersymmetry and Dynamic Magnon
Representations}, arXiv:0704.2069v2 [hep-th].
\bibitem{CDM0707}  Heng-Yu Chen, Nick Dorey, Rui F. Lima Matos, {\it Quantum Scattering of Giant
Magnons}, arXiv:0707.0668v1 [hep-th].
\bibitem{BV0707} David Berenstein, Samuel E. Vazquez, {\it Giant magnon bound states from strongly
coupled N=4 SYM}, arXiv:0707.4669v1 [hep-th].

\bibitem{Kruczenski:2004wg} M. Kruczenski, {\it Spiky strings and single trace operators in gauge
theories}, JHEP 0508 (2005) 014 [arXiv:hep-th/0410226v2].
\bibitem{Ryang:2005yd} S. Ryang, {\it Wound and Rotating Strings in
$AdS_5\times S^5$}, JHEP 0508 (2005) 047 [arXiv:hep-th/0503239v1].
\bibitem{McLoughlin:2006tz} T.~McLoughlin and X.~Wu, {\it Kinky strings in $AdS_5\times S^5$},
JHEP {\bf 0608}, 063 (2006) [arXiv:hep-th/0604193v2].
\bibitem{Chu:2006ae} Chong-Sun Chu, George Georgiou, Valentin V. Khoze,
{\it Magnons, Classical Strings and beta-Deformations}, JHEP 0611
(2006) 093, [arXiv:hep-th/0606220v2].
\bibitem{Bobev:2006fg} N.P. Bobev, R.C. Rashkov, {\it Multispin Giant
Magnons}, Phys. Rev. {\bf D 74} (2006) 046011,
[arXiv:hep-th/0607018v3].
\bibitem{IK0705}  Riei Ishizeki, Martin Kruczenski, {\it Single spike solutions for strings on S2 and
S3}, arXiv:0705.2429v1 [hep-th].
\bibitem{Mosaffa:2007} A. E. Mosaffa, B. Safarzadeh, {\it  Dual Spikes; New Spiky String
Solutions},  arXiv:0705.3131v2 [hep-th].
\bibitem{BR0706}  N.P. Bobev, R.C. Rashkov, {\it Spiky Strings, Giant Magnons and beta - deformations},
arXiv:0706.0442v2 [hep-th].
\bibitem{MNR} P. Bozhilov, {\it Neumann and Neumann-Rosochatius integrable systems from membranes on
$AdS_4\times S^7$}, arXiv:0704.3082v3 [hep-th].
\bibitem{bo-ra:0607} P. Bozhilov, R.C. Rashkov, {\it Magnon-like dispersion relation from
M-theory}, Nucl. Phys. {\bf B 768} [PM] (2007) 193-208
[arXiv:hep-th/0607116v3].

\end{thebibliography}
\end{document}